# Multi-cycle Cyclostationary based Spectrum Sensing Algorithm for OFDM Signals with Noise Uncertainty in Cognitive Radio Networks

Tadilo Endeshaw Bogale and Luc Vandendorpe
ICTEAM Institute
Universitè catholique de Louvain
Place du Levant, 2, B-1348, Louvain La Neuve, Belgium
Email: {tadilo.bogale, luc.vandendorpe}@uclouvain.be

***Abstract*— This paper proposes a simple multi-cycle cyclostationary based signal detection (spectrum sensing) algorithm for Orthogonal Frequency Division Multiplexed (OFDM) signals in cognitive radio networks. We assume that the noise samples are independent and identically distributed (i.i.d) random variables all with unknown (imperfect) variance. Our detection algorithm employ the following three steps. First, we formulate the test statistics as a ratio of two quadratic cyclic autocorrelation functions. Second, we derive a closed form expression for the false alarm probability. Third, we evaluate the detection probability of our algorithm for a given false alarm probability. The theoretical probability of false alarm expression matches with that of the simulation result. Moreover, we have observed that the proposed multi-cycle algorithm exhibits significantly superior probability of detection compared to the existing low complexity cyclostationary based and the well known energy detection algorithms.**

## I. Introduction

The multimedia wireless devices require high data rate technologies. These technologies utilize significant amount of radio spectrum. However, the available spectrum is limited. It is well known that the current wireless networks employ fixed spectrum access strategy. Given the limitations of the available spectrum, regulatory bodies such as Federal Communications Commission (FCC) have also found that the fixed spectrum access strategy utilizes the available frequency bands inefficiently [1], [2]. A cognitive radio (CR) is a promising approach for addressing the drawbacks of the current fixed spectrum access strategy. A CR is a transceiver device that can adapt its transmission parameters based on the knowledge of its surrounding radio environment. Thus, a CR network is a network that employs spectrum-aware communication protocol [1].

One of the key characteristics of a CR is its ability to detect the presence (absence) of a licensed (primary) user from its radio environment. This is performed by the signal detection (spectrum sensing) part of the CR network. The most common signal detection methods for CR network are matched filter, energy and cyclostationary based detection methods. If the characteristics of the primary user such as modulation scheme, pulse shaping filter and packet format are known perfectly, matched filter is the optimal signal detection method as it maximizes the received Signal-to-Noise ratio (SNR). In practice this information can be known a priory. The main drawback of matched filter detector is that it needs dedicated receiver to detect each signal characteristics of a primary user [3]. The energy detector does not need any information about the primary user and it is simple to implement. However, energy detector is very sensitive to noise variance uncertainty and it performs poorly whenever there is an adjacent channel interference [3], [4]. Moreover, in uncertain noise variance scenario, there is an SNR wall below which the energy detector is not able to guarantee a certain detection performance [5]. Cyclostationary based detection method is robust against noise variance uncertainty and it can reject the effect of adjacent channel interference [4]. Due to these two key advantages, cyclostationary based detection method receives a lot of attention in CR networks.

There are dozens of work on cyclostationary based signal detection algorithms (for example see [6]–[11]). Out of these papers, we are interested in the low complexity cyclostationarity based detection algorithm under noise variance uncertainty of [11]. The authors of this paper exploit the cyclostationarity behavior of Orthogonal Frequency Division Multiplexed (OFDM) signals which is induced by the cyclic prefix (CP) operation. This paper proposes a simple ratio of two autocorrelation functions (at different time lags) to test the presence of single-cycle cyclostationary based OFDM signals.

In the current paper, we propose a simple multi-cycle cyclostationary based OFDM signal detection algorithm for CR networks. We assume that the noise samples are independent and identically distributed (i.i.d) random variables all with unknown (imperfect) variance. The proposed detection algorithm employs the following steps. First, we formulate the test statistics as a ratio of two quadratic cyclic autocorrelation (CAC) functions. Second, we derive a closed form expression for the false alarm probability. Third, we evaluate the detection probability of our algorithm for a given false alarm probability. The probability of false alarm expression fits to that of the simulation result. Moreover, we have observed that the proposed multi-cycle cyclostationary based algorithm yields significantly superior probability of detection compared to the cyclostationary based detection algorithm of [11] and energy detection algorithm of [12].

The remaining part of this paper is organized as follows. Section II discusses the hypothesis test problem. In Section III, an overview of cyclostaionary based signal detection is presented. Section IV presents the proposed multi-cycle cyclostationary based detector algorithm. In Section V, computer simulations are used to compare the performance of our multi-cycle detector to that of the existing low complexity cyclostationary based and the well known energy detectors. Finally, conclusions are drawn in Section VI.

## II. Problem formulation

Let $\mathbf{x} = \{x[n]\}_{n=0}^{N-1}$ denote the observed discrete time baseband equivalent signal vector at the receiver. The observed signal has the following form [12]

$$x[n] = \begin{cases} s[n] + w[n], & H_1 \\ w[n], \quad n = 0, 1, \cdots, N-1 & H_0 \end{cases} \tag{1}$$

where $s[n]$, $w[n]$ and $N$ are the transmitted signal sample, noise sample and number of samples, respectively. The noise samples $\{w[n]\}_{n=0}^{N-1}$ are assumed to be zero-mean i.i.d random variables[1]. The aim of a CR spectrum sensing is to detect the presence or absence of the transmitted signal $s[n]$ (hereafter it is referred as a primary user signal). Hence the CR spectrum sensing turns to a binary hypothesis testing problem of $H_0$ and $H_1$.

## III. An overview of cyclostationary based signal detection

The time varying autocorrelation function of a continuous time random signal $y(t)$ is expressed as

$$R_y(t,\tau) = \mathrm{E}\{y(t)y(t+\tau)^\star\} \tag{2}$$

where $\mathrm{E}(.)$ and $(.)^\star$ denote the expectation and complex conjugate operators, respectively and $\tau$ is the time lag. Assuming that the Fourier series expansion of $R_y(t,\tau)$ is convergent, we can reexpress (2) as

$$R_y(t,\tau) = \sum_\alpha R_y^\alpha(\tau)e^{j2\pi\alpha t} \tag{3}$$

where $R_y^\alpha(\tau)$ is the Fourier coefficient which is given by

$$R_y^\alpha(\tau) = \lim_{T\to\infty}\frac{1}{T}\int_{-\frac{T}{2}}^{\frac{T}{2}} R_y(t,\tau)e^{-j2\pi\alpha t}dt. \tag{4}$$

This expression is termed as CAC function where $\alpha$ is a cyclic frequency.

For a discrete time random signal $\{x[n]\}_{n=0}^{N-1}$, the time varying autocorrelation and CAC functions can be computed as[2]

$$\begin{aligned} R_x(n,\tau) =& \mathrm{E}\{x[n]x[n+\tau]^\star\} \\ R_x^\alpha(\tau) =& \frac{1}{N}\sum_{n=0}^{N-1}\mathrm{E}\{x[n]x[n+\tau]^\star\}e^{-j2\pi\alpha n}. \end{aligned}$$

[1]In the case of nonzero mean received signal samples $\{x[n]\}_{n=0}^{N-1}$, one can remove the mean from the received samples.

[2]We would like to mention here that for discrete time random signal $\tau$ is expressed in number of samples.

A discrete time random signal $\{x[n]\}_{n=0}^{N-1}$ is said to be cyclostationary if its time varying autocorrelation function $R_x(n,\tau)$ is periodic in time, i.e., if there exists at least one $\alpha \neq 0$ with $R_x^\alpha(\tau) \neq 0$.

One of the most distinct behavior of practical communication signals is cyclostationarity [13], [14]. For communication signals, the exact $\alpha$ and $\tau$ for which $R_x^\alpha(\tau) \neq 0$ depends on different parameters like modulation scheme, symbol period and so on. For example, an OFDM signal has nonzero CAC values at

$$\begin{aligned} \tau =& N_{FFT} \\ \alpha_k =& k/(N_{FFT}+N_{CP}), k = 0, 1, -1, 2, -2, \cdots \end{aligned} \tag{5}$$

where $N_{FFT}$ and $N_{CP}$ are the lengths of FFT (number of sub-carriers) and cyclic prefix of the OFDM symbol, respectively [15], [16].

Since a discrete noise signal $\{w[n]\}_{n=0}^{N-1}$ is purely stationary (for example white Gaussian noise), $R_w^\alpha(\tau) = 0$ for all $\alpha \neq 0$. Thus, a binary hypothesis test of (1) turns to a problem of examining whether $R_x^\alpha(\tau)$ is zero (small value) or not for appropriately selected $\alpha$ and $\tau$.

In practice since finite number of samples are available, the true CAC function is replaces by its estimate i.e.,

$$\hat{R}_x^\alpha(\tau) = \frac{1}{N_\tau}\sum_{n=0}^{N_\tau-1} x[n]x[n+\tau]^\star e^{-j2\pi\alpha n}$$

where $N_\tau = N - \tau$.

## IV. Proposed multi-cycle detector

As we have explained in Section III, $R_x^\alpha(\tau)$ is the Fourier coefficient of $R_x(n,\tau)$. As can be seen from Fig. 1 of [16], when $\tau = N_{FFT}$, the ideal time varying autocorrelation function $R_x(n,\tau)$ of an OFDM signal (assuming that equal power is utilized across all sub-carriers) will incorporate a square shape. From fundamental calculus we know that the Fourier coefficients of a square signal are non-negative. Due to this reason, we propose the following simple multi-cycle cyclostationarity test statistics:

$$T_{mc} = \frac{\frac{1}{N_\tau}\left|\sum_{n=0}^{N_\tau-1}\frac{1}{\eta_n}x[n]x[n+\tau]^\star\sum_{k=0}^{K-1}e^{-j2\pi\alpha_k n}\right|^2}{\frac{1}{N_{\bar{\tau}}}\left|\sum_{n=0}^{N_{\bar{\tau}}-1}x[n]x[n+\bar{\tau}]^\star e^{-j2\pi\beta n}\right|^2} \tag{6}$$

where $\tau$ and $\{\alpha_k\}_{0=1}^{K-1}$ are the time lag and set of cyclic frequencies defined in (5), $\beta$ is any arbitrary value, $\bar{\tau} \neq N_{FFT}$, $\eta_n = \sqrt{(\sum_{k=0}^{K-1}\cos(2\pi\alpha_k n))^2 + (\sum_{k=0}^{K-1}\sin(2\pi\alpha_k n))^2}, \forall n$ and $K$ is the number of tested cyclic-frequencies[3]. Note that in (6), the denominator is incorporated just to remove the effect of noise variance uncertainty [11].

Using the above test statistics, we decide $\{x[n]\}_{n=0}^{N-1}$ of (1) as $H_0$ if $T_{mc} < \lambda$ and as $H_1$ if $T_{mc} \geq \lambda$, where $\lambda$ is a threshold value that is chosen to guarantee a certain performance. In general, $\lambda$ is selected such that the test statistics

[3]When K=1, the proposed multi-cycle detector turns to single-cycle detector.

(6) can guarantee a constant probability of detection ($P_d$) or probability of false alarm ($P_f$). In this paper, we choose the threshold $\lambda$ to guarantee a constant $P_f$. Mathematically $P_f(\lambda)$ is computed as

$$P_f(\lambda) = Pr\{T_{mc} > \lambda | H_0\} \tag{7}$$

where $Pr(.)$ denotes the probability operator. To determine $P_f(\lambda)$, the cumulative distribution function (CDF) of $T_{mc}$ under $H_0$ needs to be computed. In this regard, we examine the following Lemma:

*Lemma 1*: When $\{w[n]\}_{n=0}^{N-1}$ are zero-mean i.i.d random variables, the CDF of $T_{mc}$ under $H_0$ hypothesis can be expressed as

$$F_{T_{mc}|H_0}(r) = \frac{r}{r+1} \tag{8}$$

*Proof:* See Appendix A. ■

By applying *Lemma 1* on (7), the $P_f(\lambda)$ can thus be expressed as

$$P_f(\lambda) = Pr\{T_{mc} > \lambda | H_0\} = 1 - F_{T_{mc}|H_0}(\lambda) = 1 - \frac{\lambda}{\lambda+1} = \frac{1}{\lambda+1}. \tag{9}$$

We would like to mention here that a ratio test similar to (6) has been proposed in [17] for the detection of Worldwide Interoperability for Microwave Access (WiMAX) signal. However, as can be seen from Fig. 5 of [17], the gap between the theoretical and simulation results of the false alarm probability is large. For example, to achieve a 0.1 false alarm probability the theoretical threshold is 6.314 (from (23) of [17]) whereas the simulation threshold is around 3.6. Thus, it is not clear from [17] how we can set a threshold to ensure a constant $P_f$ for any type of noise (interference plus noise) signal and parameter settings. On the other hand, the algorithm of [17] is a single-cycle detection algorithm.

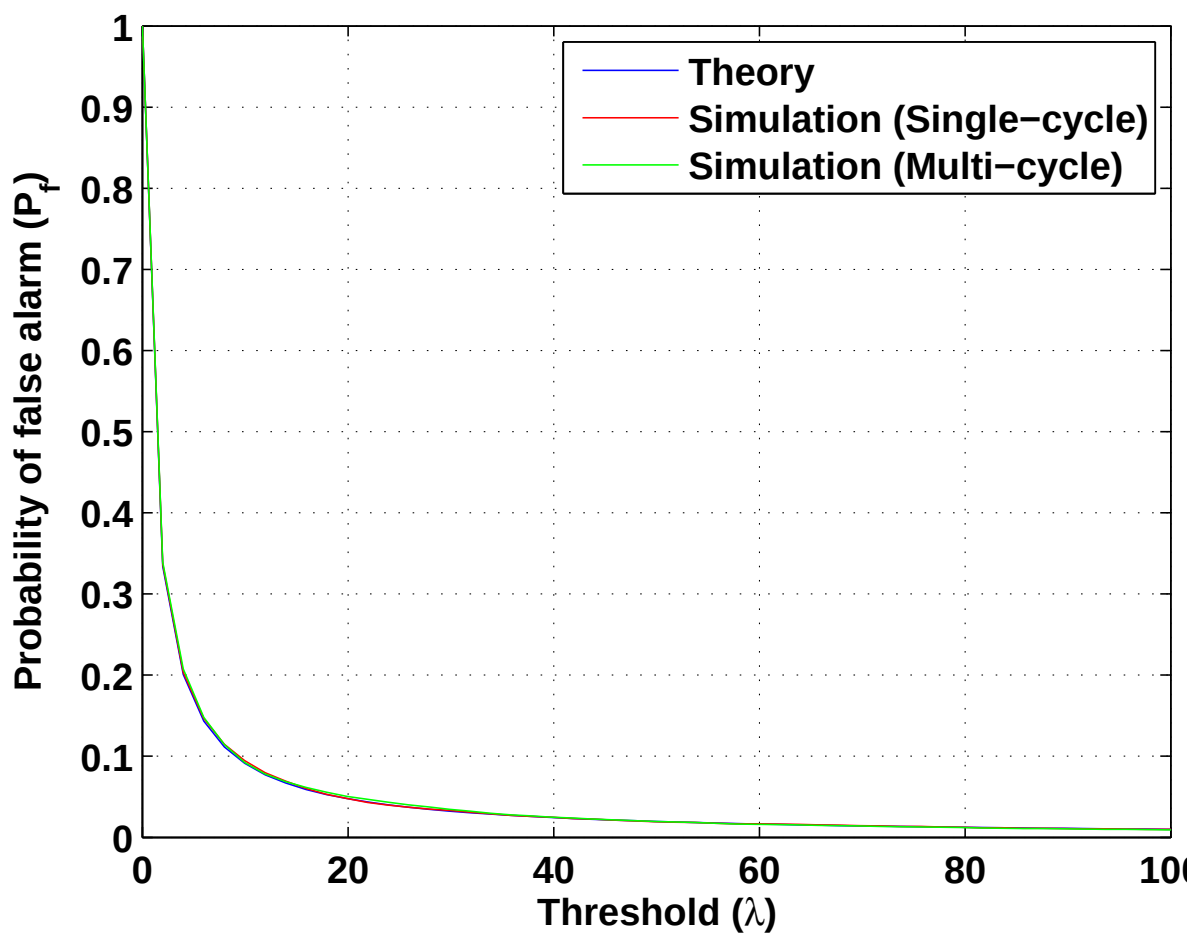

Fig. 1. Probability of false alarm for ZMCSCG noise signal.

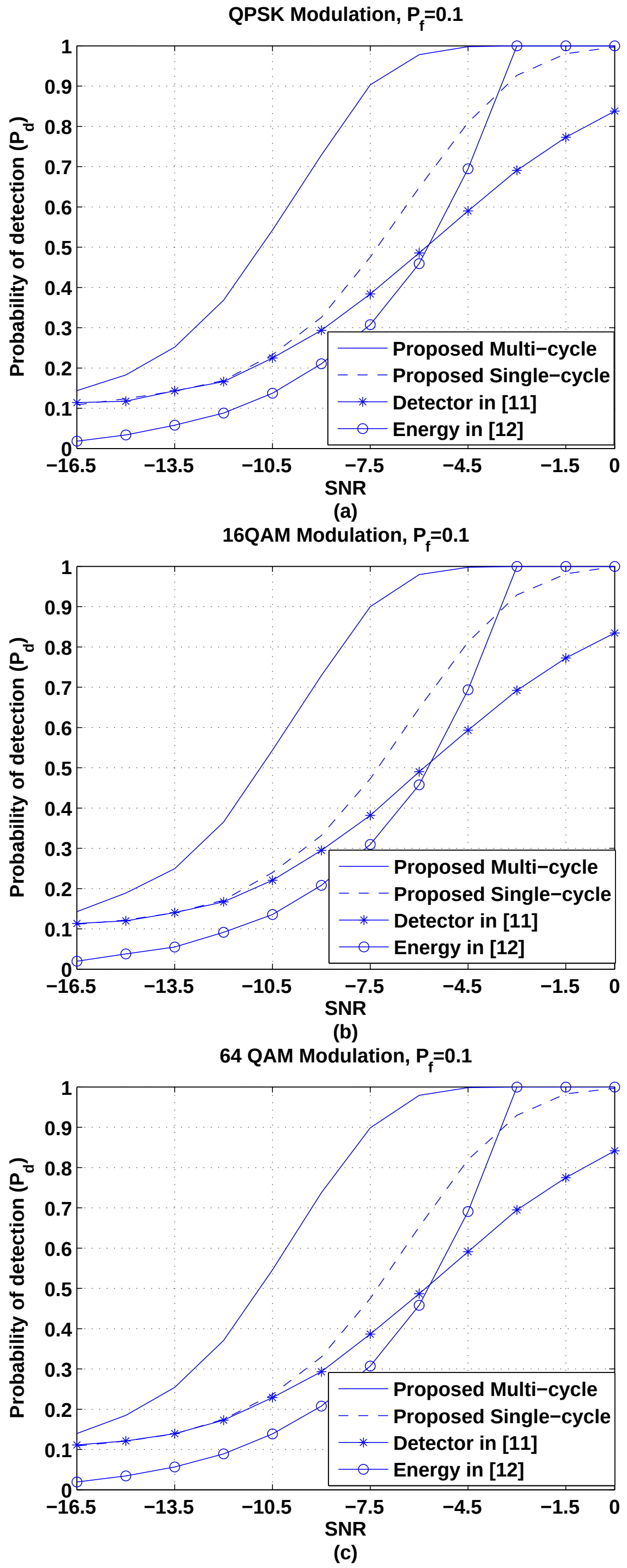

Fig. 2. Comparison of the proposed detector with the single-cycle detector in [11] and the energy detector in [12]. (a) QPSK modulation, (b) 16QAM modulation and (c) 64QAM modulation.

TABLE I

SIMULATION PARAMETERS OF WIMAX SIGNAL

| Parameter | Value |
|---|---|
| Channel BW | 5 MHz |
| FFT size ($N_{FFT}$) | 512 |
| Subcarrier spacing | 10.9KHz |
| Useful symbol duration | 91.4 $\mu$s |
| Used subcarrier index | {-240 to 1 & 1 to 240} |
| Cyclic prefix (CP) ratio | 1/8 |
| Modulation per OFDM symbol | QPSK, 16, 64 QAM |
| Number of OFDM symbols ($N_{OFDM}$) | 32 |

## V. SIMULATION RESULTS

In all of our simulation results, we assume that the noise samples are i.i.d zero mean circularly symmetric complex Gaussian (ZMCSCG) random variables all with the same variance which is not known perfectly. All results are obtained by averaging 20000 experiments. For these noise samples $R_w^\alpha(\tau) = 0, \forall\alpha$ with $\tau \neq 0$. We use $\alpha = 0$ for single-cycle results (i.e., $K = 1$), $\alpha_k, k = [-2, -1, 0, 1, 2]$ for multi-cycle results (i.e., $K = 5$) and $\beta = 0.412N$ with $\beta \neq \alpha_k, \forall k$, where $N$ is the total number of samples. According to [5], in an uncertain noise variance signal detection algorithm, the actual noise variance can be modeled as a bounded interval of $[\frac{1}{\epsilon}\sigma^2 \;\; \epsilon\sigma^2]$ for some $\epsilon = 10^{\Delta\sigma^2/10} > 1$, where the uncertainty $\Delta\sigma^2$ is expressed in dB. We assume that this bound follows a uniform distribution, i.e., $\mathcal{U}[\frac{1}{\epsilon}\sigma^2 \;\; \epsilon\sigma^2]$. The SNR is defined as $SNR \triangleq \sigma_s^2/\sigma^2$, where $\sigma_s^2$ is the variance of $\{s[n]\}_{n=0}^{N-1}$. The noise variance is the same for one observation (since it has a short duration) and follow a uniform distribution during several observations. For Fig. 2 - Fig. 4, the probability of false alarm is set to 0.1, $\tau = N_{FFT}$ and $\bar{\tau} = N_{FFT} - 2$, where $N_{FFT}$ is as given in Table I. In this table the CP ratio is defined as $N_{CP}/N_{FFT}$.

In the following discussions, first we confirm the theoretical probability of false alarm expression by numerical simulation then we compare the performance of our detector with the existing low complexity cyclostationary based and energy detectors. Finally, the effects of the cyclic prefix ratio and number of OFDM symbols on the performance of the proposed detector is presented.

In the first simulation, we verify the probability of false alarm expression of (9) by computer simulation. We take $N = 1152$ complex noise samples, $\tau = 128$ and $\bar{\tau} = 126$. Fig. 1 shows the theoretical and simulated probability of false alarm for our single-cycle and multi-cycle detection algorithms. As can be seen from this figure, the simulated $P_f(\lambda)$ matches exactly the theoretical $P_f(\lambda)$ for both single and multi-cycle frequency scenarios.

In the second simulation, we compare the performance of the proposed detector with the single-cycle detector of [11] and the energy detector of [12]. For the comparison, we employ a WiMAX signal with the parameters as shown in Table I [10] and $\Delta\sigma^2 = 1dB$. Fig. 2 shows that the proposed single-cycle and multi-cycle detectors outperform the single-cycle detector of [11] and the energy detector of [12] for all modulation scheme (i.e., QPSK, 16 QAM and 64 QAM). Moreover, these figures show that the performance of the proposed detector does not depend on the modulation scheme of the primary user signal which is a desirable property as it can detect primary user signals that employ adaptive modulation schemes just with a single algorithm.

In the third simulation, we examine the effect of the CP ratio on the performance of cyclostationary based detectors and the effect of noise uncertainty level on the performance of energy detector. In this regard we fix the SNR to $SNR = -10dB$ and modulation scheme to QPSK. We take $\Delta\sigma^2 = [0.5dB, \; 1dB]$, the CP ratios are [1/4,1/8,1/16,1/32] and all the other settings are like in Table I. Fig. 3 shows the performance of all detectors for these settings. As we can see from Fig. 3, when we increase the CP ratio, the detection probability of the cyclostationary based detectors increase whereas as expected, due to noise uncertainty, the detection probability of energy detector is not improved.

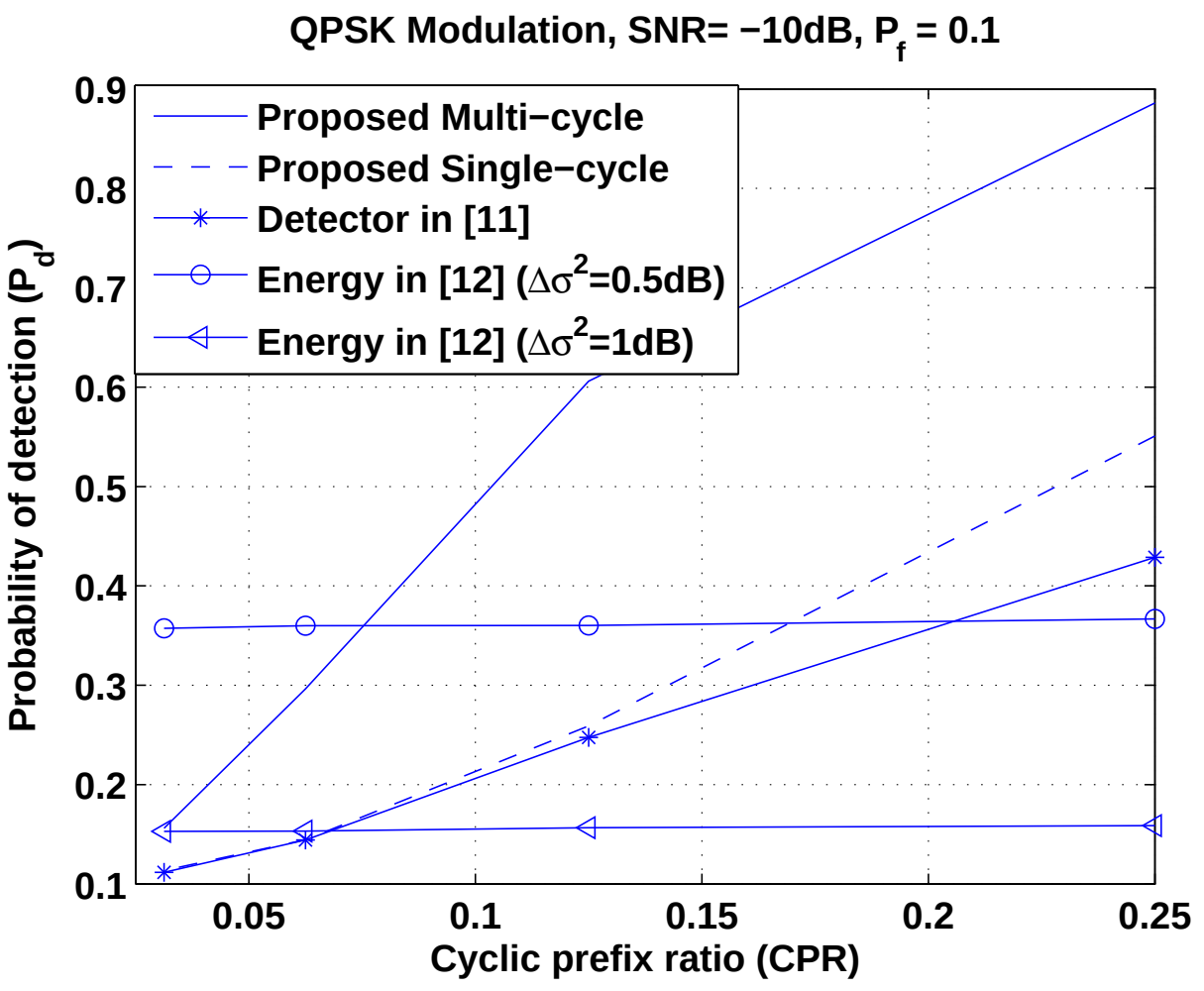

Fig. 3. Comparison of the proposed detector with the single-cycle detector in [11] and energy detector in [12] for different CP ratio.

In the last simulation, we examine the effect of the number of OFDM symbols on the performance of the detectors. For this simulation, we employ the settings of Table I with $N_{OFDM} = [16, 32, 64, 128]$. We set $SNR = -10dB$, $\Delta\sigma^2 = [0.5dB, \; 1dB]$ and modulation scheme is QPSK. Fig. 4 shows the performance of different detectors for different number of OFDM symbols. This figure shows that increasing the number of OFDM symbols improves the performance of all cyclostationary based detectors. However, due to noise uncertainty, the performance of energy detector is not improved.

As we can see from Fig. 2 - Fig. 4, in a cyclostationary based signal detector, superior performance improvement can be achieved by utilizing multiple cyclic frequencies.

## VI. CONCLUSIONS

In this paper we propose a simple multi-cycle cyclostationary based spectrum sensing algorithm for OFDM signals in CR networks. We assume that the noise samples are i.i.d random variables with unknown (imperfect) variance. For the practically relevant i.i.d ZMCSCG noise with imperfect variance scenario, we demonstrate that the proposed multi-cycle cyclostationary based spectrum sensing algorithm achieves significantly superior detection probability compared to the single-cycle cyclostationary based and the well known energy detection algorithms.

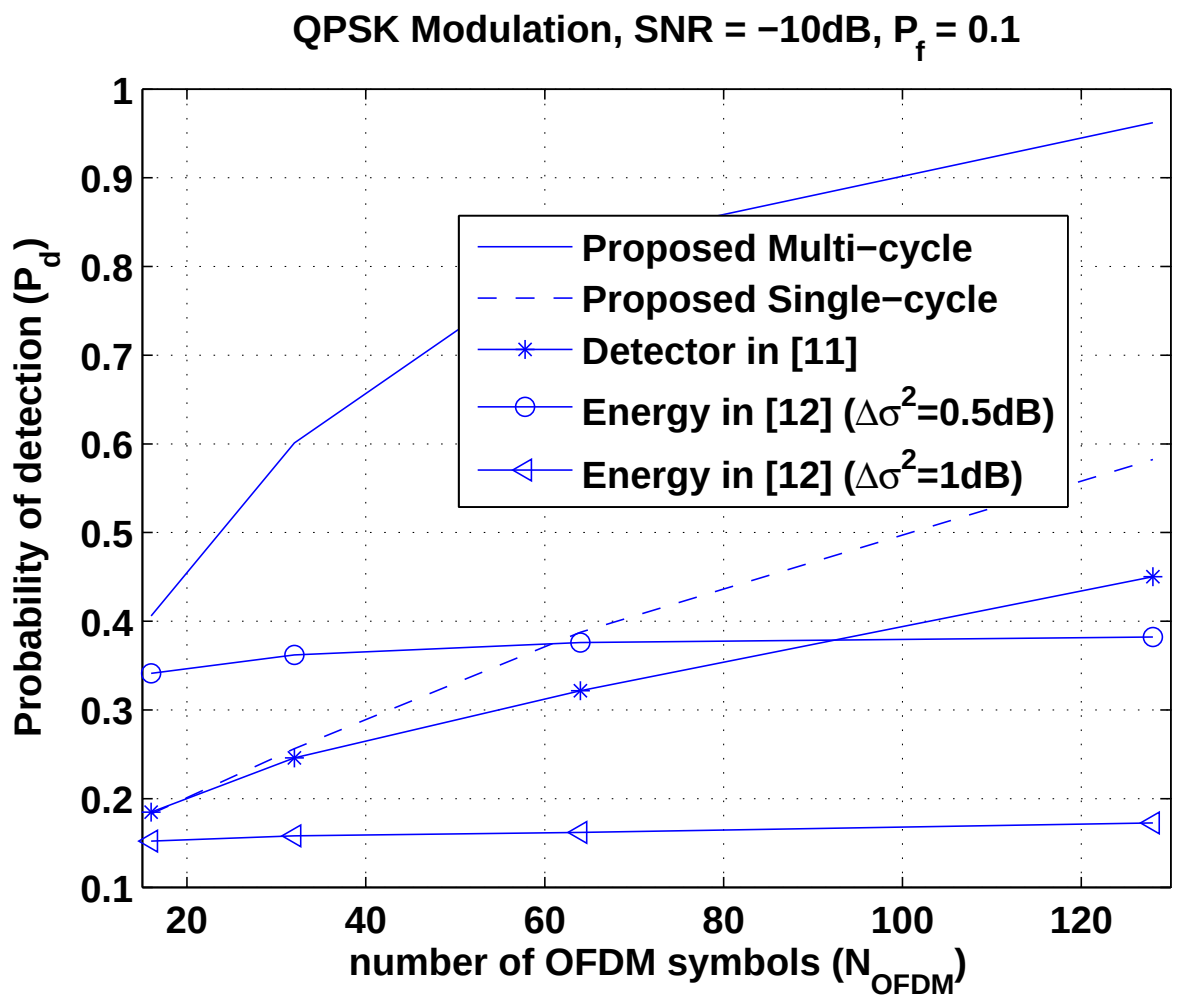

Fig. 4. Comparison of the proposed detector with the single-cycle detector in [11] and energy detector in [12] for different OFDM symbol length.

## VII. Acknowledgements

The authors would like to thank SES for the financial support, the french community of Belgium for funding the ARC SCOOP and BELSPO for funding the IAP BESTCOM project. In particular, we would like to thank Dr. Joel Grotz for his technical discussions, comments and suggestions during the work of this research.

## Appendix A
## Proof of Lemma 1

*Proof:* In this proof, under $H_0$ hypothesis, first we show that the numerator and denominator of $T_{mc}$ are independent chi-square random variables then we derive the CDF of $T_{mc}$. Under $H_0$ hypothesis, $T_{mc}$ can be expressed as

$$T_{mc} = \frac{\frac{1}{N_\tau}\left|\sum_{n=0}^{N_\tau-1}\frac{1}{\eta_n}w[n]w[n+\tau]^\star\sum_{k=0}^{K-1}e^{-j2\pi\alpha_k n}\right|^2}{\frac{1}{N_{\bar\tau}}\left|\sum_{n=0}^{N_{\bar\tau}-1}w[n]w[n+\bar\tau]^\star e^{-j2\pi\beta n}\right|^2} \triangleq \frac{A^2}{B^2}.$$

By decomposing $w[n]$ as $w[n] = w_r[n]+jw_i[n]$, where $w_r[n]$ and $w_i[n]$ are real and imaginary parts of $w[n]$, $A^2$ and $B^2$ can be rewritten as

$$\begin{aligned} A^2 &= A_1^2 + A_2^2 \\ B^2 &= B_1^2 + B_2^2 \end{aligned} \tag{10}$$

where

$$A_1 = \frac{1}{\sqrt{N_\tau}}\sum_{n=0}^{N_\tau-1}\frac{1}{\eta_n}[(w_r[n]w_r[n+\tau]+w_i[n]w_i[n+\tau])f_n - (w_r[n]w_i[n+\tau]-w_i[n]w_r[n+\tau])g_n]$$

$$A_2 = \frac{1}{\sqrt{N_\tau}}\sum_{n=0}^{N_\tau-1}\frac{1}{\eta_n}[(w_r[n]w_r[n+\tau]+w_i[n]w_i[n+\tau])g_n - (w_r[n]w_i[n+\tau]-w_i[n]w_r[n+\tau])f_n]$$

$$B_1 = \frac{1}{\sqrt{N_{\bar\tau}}}\sum_{n=0}^{N_{\bar\tau}-1}[(w_r[n]w_r[n+\bar\tau]+w_i[n]w_i[n+\bar\tau])\cos(2\pi\beta n) - (w_r[n]w_i[n+\bar\tau]-w_i[n]w_r[n+\bar\tau])\sin(2\pi\beta n)]$$

$$B_2 = \frac{1}{\sqrt{N_{\bar\tau}}}\sum_{n=0}^{N_{\bar\tau}-1}[(w_r[n]w_r[n+\bar\tau]+w_i[n]w_i[n+\bar\tau])\sin(2\pi\beta n) - (w_r[n]w_i[n+\bar\tau]-w_i[n]w_r[n+\bar\tau])\cos(2\pi\beta n)]$$

$f_n = \sum_{k=0}^{K-1}\cos(2\pi\alpha_k n)$ and $g_n = \sum_{k=0}^{K-1}\sin(2\pi\alpha_k n)$.

Assuming that $N$ is sufficiently large (which is a reasonable assumption in CR), we can apply central limit theorem for $A_1, A_2, B_1$ and $B_2$ [15]. Thus, $A_1, A_2, B_1$ and $B_2$ are zero mean Gaussian random variables with variances given by

$$\sigma_{A_1}^2 = \sigma_{A_2}^2 = \frac{1}{N_\tau}\sum_{n=0}^{N_\tau-1}\sigma^4\frac{f_n^2+g_n^2}{\eta_n^2} = \sigma^4$$

$$\sigma_{B_1}^2 = \sigma_{B_2}^2 = \frac{1}{N_{\bar\tau}}\sum_{n=0}^{N_{\bar\tau}-1}\sigma^4 = \sigma^4$$

where $\frac{\sigma^2}{\sqrt{2}} = \mathrm{E}\{w_r[n]^2\} = \mathrm{E}\{w_i[n]^2\}, \forall n$. Furthermore, for $\bar\tau \neq \tau$, $A_1, A_2, B_1$ and $B_2$ are uncorrelated and hence are independent. Since $\mathcal{N}(0,\sigma^4)$ is equivalently expressed as $\sigma^2\mathcal{N}(0,1)$, we can rewrite $T_{mc}$ as

$$T_{mc} = \frac{A^2}{B^2} = \frac{A_1^2+A_2^2}{B_1^2+2^2} = \frac{\tilde{A}_1^2+\tilde{A}_2^2}{\tilde{B}_1^2+\tilde{B}_2^2} \triangleq \frac{\tilde{A}^2}{\tilde{B}^2}$$

where $\tilde{A}_1, \tilde{A}_2, \tilde{B}_1$ and $\tilde{B}_2$ are zero mean independent Gaussian random variables all with unit variance (i.e., $T_{mc}$ does not depend on the actual variance of $\{w[n]\}_{n=0}^{N-1}$).

Therefore, $\tilde{A}^2$ and $\tilde{B}^2$ are independent chi-square distributed RVs each with 2 degrees of freedom. Consequently, according to [18], $T_{mc}$ is an F-distributed random variable with CDF given by

$$F(r,d_1,d_2) = I_{\frac{d_1 r}{d_1 r+d_2}}(d_1/2, d_2/2) \tag{11}$$

where $d_1 = d_2 = 2$ and

$$\begin{aligned} I_z(a,b) &= J(z;a,b)/J(a,b) \\ J(z;a,b) &= \int_0^z t^{a-1}(1-t)^{b-1}dt \\ J(a,b) &= \int_0^1 t^{a-1}(1-t)^{b-1}dt. \end{aligned} \tag{12}$$

Substituting $a = b = \frac{d_1}{2} = 1, z = \frac{1}{r+1}$ in 12

Solving (11) gives (8). ■